\documentclass[journal]{IEEEtran}
\usepackage{graphics}
\usepackage{colortbl}
\usepackage{multicol}
\usepackage{diagbox}
\usepackage{lipsum}
\usepackage{color}
\usepackage[cmex10]{amsmath}
\usepackage{amsthm}
\usepackage{amssymb}
\usepackage{mathrsfs}
\usepackage{mathtools}
\usepackage{amsbsy}
\usepackage[colorlinks=true,bookmarks=false,citecolor=blue,urlcolor=blue]{hyperref}
\usepackage{xcolor}
\usepackage{mathrsfs}
\usepackage{setspace}
\usepackage{float}
\usepackage{graphicx}
\usepackage{pstool}
\usepackage{epstopdf}
\usepackage{lettrine}
\usepackage{psfrag}
\usepackage{newclude}
\usepackage[normalem]{ulem}
\usepackage{latexsym}
\usepackage{algpseudocode}
\usepackage{algorithm,algpseudocode}
\usepackage{algorithmicx}
\usepackage{gensymb}
\usepackage{marginnote}
\usepackage{multirow}
\usepackage{amsmath}
\usepackage{amsmath,bm}
\usepackage{wasysym}
\usepackage{mathrsfs}
\usepackage{amsmath,cite,amsfonts,amssymb,color}
\usepackage[T1]{fontenc}
\usepackage{graphicx,psfrag}
\usepackage{pstool}
\usepackage{algorithm}
\usepackage{algpseudocode}
\usepackage{comment}
\usepackage{tabularx}
\usepackage{caption}
\usepackage{csquotes}
\usepackage{float}
\usepackage{lettrine}
\usepackage{makecell}
\usepackage{textcomp, gensymb}
\usepackage{relsize}

\ifCLASSOPTIONcompsoc
\usepackage[caption=false,font=normalsize,labelfon
t=sf,textfont=sf]{subfig}
\else
\usepackage[caption=false,font=footnotesize]{subfig}
\fi

\allowdisplaybreaks

\graphicspath{{figures/}}
\allowdisplaybreaks

\newcommand{\RNum}[1]{\uppercase\expandafter{\romannumeral #1\relax}}

\usepackage{mathtools}

\begin{document}
\title{Secrecy Rate Analysis of STAR-RIS in Presence of Energy Harvesting Eavesdroppers}
\author{\IEEEauthorblockN{Mohammad Reza Kavianinia\thanks{Mohammad Reza Kavianinia and Mohammad Javad Emadi are with the Department of Electrical Engineering, Amirkabir University of Technology (Tehran Polytechnic), Tehran, Iran (E-mails: \{mrezakaviani,mjemadi\}@aut.ac.ir).} and Mohammad Javad Emadi
}}   
\maketitle
\begin{abstract}
This work peruses a simultaneous transmitting and reflecting reconfigurable intelligent surface (STAR-RIS) based wireless system wherein untrusted energy harvesting nodes wiretap the legitimate signal. The achievable secrecy rate and the harvested energy for multicast communication are computed as functions of STAR-RIS transmission and reflection coefficients (TARCs). Thus, to maximize the secrecy rate while satisfying the minimum required harvested energies, the non-convex optimization problem is investigated. By using the Dinkelbach transformation and semi-definite relaxation technique, the optimization problem is transformed into a convex problem. Eventually, the numerical results are reported to specify the effects of the size of the STAR-RIS and optimizing its TARCs elements on magnifying the secrecy rate-energy harvesting region.
\end{abstract}
\begin{IEEEkeywords}
 Simultaneous transmitting and reflecting reconfigurable intelligent surface, Secrecy, Multicast communication, Energy harvesting.
\end{IEEEkeywords}
\section{Introduction}
Intelligent reflecting surfaces (RISs) captured research consideration of future sixth-generation (6G) for boosting the performance of wireless communication systems \cite{renzo2019smart,wu2019towards}. Lately, a novel concept of simultaneous transmitting and reflecting RISs (STAR-RISs) has been proposed, which can simultaneously transmit and reflect the incident signals, which leads to half-space into full-space coverage conversion \cite{liu2021star}. Therefore, by adjusting the transmitted and reflected signals by a STAR-RIS element through its corresponding transmission and reflection coefficients (TARCs), the signal propagation is going to be controllable.
From another point of view, for the sake of pervasive usage of sensor networks and the significance of green communications, wireless power transfer and energy harvesting  techniques have achieved many profits to extend the lifetime and efficiency of the networks \cite{ng2019wireless}. In addition, the simultaneous wireless information and power transfer (SWIPT) scheme has facilitated a receiver to not only harvest energy but also receive information \cite{abedi2018power,tang2019joint}. In \cite{masoumi2019performance}, the performance of a power splitting-based SWIPT RIS-based system is evaluated with a traditional decode and forward relaying one. It is indicated that by increasing the number of reflecting elements of the RIS, the system outperforms the conventional relaying system. Furthermore, investigating the conventional green communication systems, considering the physical layer secrecy in presence of energy harvesting nodes are of profits \cite{tang2018optimization}. In \cite{tang2018optimization}, a SWIPT-enable non-orthogonal multiple access systems in presence of energy harvesting
\begin{figure}[H]
    \centering
    \pstool[scale=0.5]{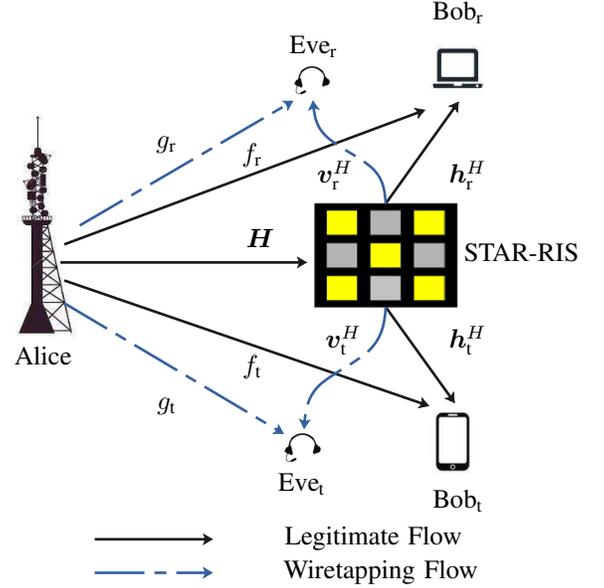}{
    \psfrag{A}{\hspace{-0.2cm}$\text{Eve}_\text{t}$}
    \psfrag{B}{\hspace{-0.35cm} $\text{Bob}_\text{t}$}
    \psfrag{C}{\hspace{-0.2cm}$\text{Eve}_\text{r}$}
    \psfrag{D}{\hspace{-0.25cm} $\text{Bob}_\text{r}$}
    \psfrag{E}{\hspace{-0.55cm} \text{STAR-RIS}}
    \psfrag{F}{\hspace{-0.3cm} \text{Alice}}
    \psfrag{G}{${g}_\text{r}$}
    \psfrag{H}{${f}_\text{r}$}
    \psfrag{I}{$\boldsymbol{H}$}
    \psfrag{J}{${f}_\text{t}$}
    \psfrag{K}{${g}_\text{t}$}
    \psfrag{L}{\hspace{-0.2cm}$\boldsymbol{v}_\text{r}^H$}
    \psfrag{M}{$\boldsymbol{h}_\text{r}^H$}
    \psfrag{N}{\hspace{-0.2cm} $\boldsymbol{v}_\text{t}^H$}
    \psfrag{O}{$\boldsymbol{h}_\text{t}^H$}
    \psfrag{P}{\text{Legitimate Flow}}
    \psfrag{Q}{\text{Wiretapping Flow}}
    }
    \caption{STAR-RIS-aided secure wireless communication system in presence of an energy harvesting eavesdroppers.}
    \label{fig:1}
\end{figure}
eavesdroppers is investigated and by optimizing the power allocation, the sum secrecy rates are maximized subject to minimum harvested energy at the eavesdroppers. In \cite{saeidi2020irs}, RIS-based secrecy rate analysis in presence of an energy harvesting eavesdropper is considered and the improvement of the secrecy rate while charging the battery of the energy harvesting eavesdropper wirelessly, is studied. Therefore, surveying the potential usage of the RIS and SWIPT scheme to boost physical layer secrecy in sensor networks has become one of the significant research attention;
In \cite{shen2019secrecy}, an RIS-aided system is studied to maximize the secrecy rate by jointly optimizing the reflecting coefficients and beamformer at the transmitter. Improving the security in a multiple-input single-output network to exploit the potential of STAR-RIS is considered in \cite{niu2021simultaneous}. In \cite{zhang2022security}, the secure transmission of the STAR-RIS aided communication system is investigated. In this paper, we survey a wiretap channel in presence of a STAR-RIS to enhance the secrecy rate for multicast communication while charging the battery of the energy harvesting eavesdroppers in wireless mode. It is considered that Alice wants to set up secure communications with $\text{Bob}_r$ and $\text{Bob}_t$ through the STAR-RIS, and there are  battery-limited untrusted sensor nodes that may eavesdrop on the signal. In our work, we assumed the stronger eavesdropping channel with respect to the legitimate channel, while the STAR-RIS optimized TARCs have to provide the minimum required energy for the wireless sensor nodes. The secrecy rate and energy harvesting functions are computed based on STAR-RIS TARCs, and the corresponding non-convex optimization problem is discussed. Nevertheless, this optimization problem is challenging to solve. We transformed the problem into a convex and tractable one by utilizing the Dinkelbach transformation and semi-definite relaxation (SDR) technique. It is indicated that, by increasing the number of STAR-RIS elements, one can enhance the secrecy rate and/or the energy harvesting requirement.\\
\indent\textit{Organization}: The rest of this paper is organized as follows.
In Section II, the proposed system model is introduced and presents the STAR-RIS-based wiretap channel model. In Section III, the secrecy rate maximization problem subject to harvesting minimum energy at eavesdropper is formulated. Numerical results are discussed in Section IV, and
finally, Section V concludes the paper.\\
\indent\textit{Notations}: $\mathbb{C}^{M \times N}$ denotes the space of $M \times N$ complex valued matrices. For a square matrix $\boldsymbol{A}$ , $\text{tr}(\boldsymbol{A})$ denotes its trace
and $\boldsymbol{A} \succeq 0$ denotes that $\boldsymbol{A}$ is positive semi-definite matrix.
$\text{rank}(\boldsymbol{A})$ denotes the rank of $\boldsymbol{A}$. For complex valued vector $\boldsymbol{x}$, $\big|\boldsymbol{x}\big|$ denotes its euclidean norm.
\section{System Model}
We consider a four-node STAR-RIS aided multicast downlink communication system as shown in Fig. \ref{fig:1}, in which a legitimate single- antenna source, i.e. Alice, exploits an M-element STAR-RIS to serve two single-antenna receivers, which are called $\text{Bob}_\text{r}$ and $\text{Bob}_\text{t}$. For enhancing link performance, a STAR-RIS is deployed to jointly improve the secrecy rate and the harvested energy at eavesdropper nodes, i.e. $\text{Eve}_\text{r}$ and $\text{Eve}_\text{t}$. In addition, $\text{Bob}_\text{r}$-$\text{Eve}_\text{t}$ and $\text{Bob}_\text{t}$-$\text{Eve}_\text{t}$ pairs are located at the reflection and the transmission region of STAR-RIS, respectively. We consider that the full channel state information of all the channels is available at the source. It is worth noting that $\text{Eve}_\text{r}$ and $\text{Eve}_\text{t}$ are kind of energy harvesting sensor nodes with being trusted at energy level but untrusted from Alice-$\text{Bob}_\text{r}$ and  Alice-$\text{Bob}_\text{t}$ perspectives.
\subsection{Signal Model of STAR-RIS}
Assume that the STAR-RIS has $M$ elements, where $m \in \mathcal{M}  \triangleq \{ 1, 2, ..., M\}$ and $M$ denotes the total number of elements. The transmitted and reflected signals on a given element of STAR-RIS are given as $t_m =\sqrt{\beta_m^t}e^{j\phi_m^t}X_s$ and $r_m = \sqrt{\beta_m^r}e^{j\phi_m^r}X_s$, respectively, where $\sqrt{\beta_m^t} \in \left [0,1\right] $, $\phi_m^t \in \left [0,2\pi\right] $ and $\sqrt{\beta_m^r} \in \left [0,1\right]$, $\phi_m^r \in \left [0,2\pi\right]$ and $X_s$ are the amplitude and phase shift response of the $m$-th element and information symbol, respectively. As pointed at [], for an ideal STAR-RIS, $\sqrt{\beta_m^t}$ and $\sqrt{\beta_m^r}$ can be chosen independently. However the amplitude alignments for transmission and reflection (i.e., $\sqrt{\beta_m^t}$ and $\sqrt{\beta_m^r}$) are coupled by the law of energy conservation and the sum of the energies of the transmitted and reflected signals has to be equal to the energy of the incident signal, which is equal to $\beta_m^t + \beta_m^r = 1, \forall m \in \mathcal{M}$.     
\subsection{STAR-RIS's Three Practical Protocols }
STAR-RIS has three proposed practical protocols as briefly described in the following.
\subsubsection{Energy Splitting (ES)} 
In the ES protocol, all elements of STAR-RIS can work in transmitting and reflecting modes simultaneously. therefore, the TARCs are modelled as $\boldsymbol{\Phi}_{t/r}^{ES}=\textrm{diag}(\sqrt{\beta_1^{t/r}}e^{j\phi_1^{t/r}}, \dots , \sqrt{\beta_M^{t/r}}e^{j\phi_M^{t/r}})$, where $\beta_m^t,\beta_m^r \in \left[0,1\right]$, $\beta_m^t + \beta_m^r =1$, and $\phi_m^t,\phi_m^r \in \left[0,2\pi\right), \forall m\in\mathcal{M}$. 
\subsubsection{Mode Selection (MS)}
In the MS protocol, all elements of STAR-RIS are divided into two parts, so that one part with $M_t$  elements working in transmitting mode, while another part with $M_r$ elements working in the reflecting mode, as such satisfying $M_t + M_r=M$. Hence, the TARCs are modelled as $\boldsymbol{\Phi}_{t/r}^{MS}=\textrm{diag}(\sqrt{\beta_1^{t/r}}e^{j\phi_1^{t/r}}, \dots , \sqrt{\beta_M^{t/r}}e^{j\phi_M^{t/r}})$, where $\beta_m^t,\beta_m^r \in \{0,1\}, \beta_m^t + \beta_m^r =1$, and $\phi_m^t,\phi_m^r \in \left[0,2\pi\right), \forall m\in\mathcal{M}$.
\subsubsection{Time Switching (TS)} In the TS protocol, all elements are switched between transmitting and reflecting mode by STAR-RIS in various orthogonal time durations. Let $\lambda_t$ and $\lambda_r$ be the time slot for  transmitting and reflecting modes, respectively, fulfilling $\lambda_t + \lambda_r=1 $. Therefore, the TARCs are given by $\boldsymbol{\Phi}_{t/r}^{TS}=\textrm{diag}(e^{j\phi_1^{t/r}}, \dots , e^{j{\phi_M^{t/r}}})$, where $\phi_m^t,\phi_m^r \in \left[0,2\pi\right), \forall m \in \mathcal{M}$. 
\subsection{Signal Transmissions and Receptions}
Signal transmissions and receptions at each node are studied in the following.
\begin{itemize}
    \item Alice sends the information symbol $X_s \sim \mathcal{C}\mathcal{N} (0,P_s)$ where $P_s$ is the average transmission power.
    \item The M-element STAR-RIS receives the following vector
    \begin{equation}
    \boldsymbol{Y}_{SR}= \boldsymbol{H}X_s\label{eq:1},
    \end{equation}
    where $\boldsymbol{H} \in \mathbb{C}^{M \times 1}$ denotes the channel coefficients between Alice-STAR-RIS pair. 
    \item The received signal at $\text{Bob}_k$ is given by
    \begin{equation}
    Y_{B_k}= (\boldsymbol{h}_k^H\boldsymbol{\Phi}_k\boldsymbol{H} + f_k)X_s + N_{B_k}\label{eq:2},
    \end{equation}
    where $\boldsymbol{h}_k^H \in \mathbb{C}^{1 \times M}$, $f_k \in \mathbb{C}$ and $\boldsymbol{\Phi}_k \in \mathbb{C}^{M \times M}$ denote the channel coefficients, $\forall  k \in \{r, t\}$, between STAR-RIS-$\text{Bob}_k$ pair, Alice-$\text{Bob}_k$ pair and diagonal matrix expresses amplitude and phase shift of STAR-RIS in reflecting/transmitting mode, respectively, and $N_{B_k}\sim \mathcal{C}\mathcal{N} (0,\sigma_{B_k}^2)$ models the circular symmetric complex additive white Gaussian noise (AWGN) at $\text{Bob}_k$. 
    \item The received signal at single-antenna $\text{Eve}_k$ is given by
    \begin{equation}
        Y_{E_k}= (\boldsymbol{v}_k^H\boldsymbol{\Phi}_k\boldsymbol{H} + g_k)X_s + N_{E_k}\label{eq:3},
    \end{equation}
    where $\boldsymbol{v}_k^H \in \mathbb{C}^{1 \times M}$ and $g_k \in \mathbb{C}$ denote the channel coefficients, $\forall  k \in \{r, t\}$, between STAR-RIS-$\text{Eve}_k$ pair and Alice-$\text{Eve}_k$ pair, respectively, and $N_{E_k}\sim \mathcal{C}\mathcal{N} (0,\sigma_{E_k}^2)$ models the AWGN at $\text{Eve}_k$. 
    For the simplicity and without loss of generality, we assume that $\sigma_{B_r}^2 = \sigma_{B_t}^2 = \sigma_{E_r}^2 = \sigma_{E_t}^2 = \sigma^2$. Hence, by ignoring the noise power compared to that signal, the harvested energy at the $\text{Eve}_k$, $\forall k \in \{r, t\}$,  becomes
    \begin{equation}
        P_{E_k}={\big|\boldsymbol{v}_k^H\boldsymbol{\Phi}_k\boldsymbol{H} + g_k \big|}^2P_s\label{eq:4}.
    \end{equation}
\end{itemize}
\subsection{Secrecy rate} 
By using the signal receptions given in \eqref{eq:2} and \eqref{eq:3}, the signal-to-noise ratio (SNR) at the $\text{Bob}_k$ and $\text{Eve}_k$, $k \in \{r, t\}$ can be expressed as
\begin{equation}
    \text{SNR}_{\text{Bob}_k}= \frac{{\big|\boldsymbol{h}_k^H\boldsymbol{\Phi}_k\boldsymbol{H} + f_k\big|}^2P_s}{\sigma^2}\label{eq:5},
\end{equation}
\begin{equation}
    \text{SNR}_{\text{Eve}_k}= \frac{{\big|\boldsymbol{v}_k^H\boldsymbol{\Phi}_k\boldsymbol{H} + g_k\big|}^2P_s}{\sigma^2}\label{eq:6}.
\end{equation}
Thus, by utilizing  the conventional secrecy rate analysis
for the fading channel \cite{el2011network}, we formulate the secrecy rate as a
function of STAR-IRS TARCs parameters for the ES or MS protocol as
\begin{equation}
    R_k(\boldsymbol{\Phi}_k)=\big[\ln(1+\text{SNR}_{\text{Bob}_k}) - \ln(1+\text{SNR}_{\text{Eve}_k}) \big]^+\label{eq:7},
\end{equation}
and for TS protocol the secrecy rate as a function of STAR-RIS TARCs parameters is given by
\begin{equation}
    R_k(\boldsymbol{\Phi}_k)=\lambda_k\big[\ln(1+\text{SNR}_{\text{Bob}_k}) - \ln(1+\text{SNR}_{\text{Eve}_k}) \big]^+\label{eq:8},
\end{equation}
where $\big[x\big]^+ \triangleq \max\{x, 0 \}$.
\section{PROBLEM STATEMENT AND ANALYSIS}
In the following, we propose the optimization problem, and by using the Dinkelbach algorithm, which was developed to solve convex
nonlinear fractional programming problems and it is transformed into a semi-definite problem \cite{schaible1976fractional}. Thus, we optimize the STAR-RIS's TARCs to meet the energy harvesting requirement at Eves while maximizing the secrecy rate. 
\subsection{Problem Formulation}
In this paper, we aim to optimize the TARCs, i.e. $\boldsymbol{\Phi}_r$ and $\boldsymbol{\Phi}_t$, to jointly maximize the achievable secure data rate at Bob's and harvest a minimum required energy at the Eves as well. Therefore, the corresponding optimization problem $\mathcal{P}_1$ for the ES, MS and TS is defined as
\begin{subequations}
    \begin{align}
        \begin{split}
               \mathcal{P}_{1}:\max_{\boldsymbol{\Phi}_r ,\boldsymbol{\Phi}_t}
                \quad & \mathlarger{\mathlarger{\sum}}_{k\in \{r, t\}} \ln\Bigg( \frac{\sigma^2 +{\big|\boldsymbol{h}_k^H\boldsymbol{\Phi}_k\boldsymbol{H} + f_k\big|}^2P_s}{\sigma^2 + {\big|\boldsymbol{v}_k^H\boldsymbol{\Phi}_k\boldsymbol{H} + g_k\big|}^2P_s }\Bigg) \Bigg / \\
                \max_{\substack{\boldsymbol{\Phi}_r ,\boldsymbol{\Phi}_t,\\ \lambda_r, \lambda_t}}
                \quad & \mathlarger{\mathlarger{\sum}}_{k\in \{r, t\}} \lambda_k\ln\Bigg( \frac{\sigma^2 +{\big|\boldsymbol{h}_k^H\boldsymbol{\Phi}_k\boldsymbol{H} + f_k\big|}^2P_s}{\sigma^2 + {\big|\boldsymbol{v}_k^H\boldsymbol{\Phi}_k\boldsymbol{H} + g_k\big|}^2P_s }\Bigg)\label{eq:9a}
        \end{split}\\
        \textrm{s.t.} \quad & P_{E_k} \ge E_k, \forall k \in \{r, t\},\label{eq:9b}\\
        \quad & \boldsymbol{\Phi}_k^X \in \mathcal{F}^X, \forall k \in \{r, t\},\label{eq:9c}\\
        \quad & 0 \le \lambda_r,\lambda_t \le 1, \lambda_r+\lambda_t=1, \label{eq:9d} 
    \end{align}
\end{subequations}
where \eqref{eq:9b} denotes that the harvested energy $P_{E_k}$ must be larger than the required energy $E_k$, $\forall  k \in \{r, t\}$ and $X \in \{\text{ES,MS,TS}\}$ specifies the employed STAR-RIS operating protocol, $\mathcal{F}^X$ characterizes the corresponding feasible set for the TARCs matrices, and the time allocation variables, $\lambda_k$ , and constraint \eqref{eq:9d} are only valid when the TS protocol is utilized, i.e., $X$ = TS. The first term in \eqref{eq:9a}  is referred to the ES and MS, and the second term in \eqref{eq:9a} is only valid for the TS. Since the objective function is not a concave, and there exist non-convex constraints, it is not easy and challenging to analyze and solve the optimization problem $\mathcal{P}_{1}$. Therefore, in the following, by transforming the optimization problem $\mathcal{P}_{1}$, we can analyze the equivalent optimization problem by a standard solver such as CVX \cite{grant2014cvx}.
\subsection{Problem Transformation} 
for sake of solving the problem $\mathcal{P}_{1}$, it is transformed into the following form. Defining $\boldsymbol{W}^k \in \mathbb{C}^{(M+1) \times 1}$ and $\boldsymbol{G}^k \in \mathbb{C}^{(M+1) \times 1}$ as follows,
\begin{equation}
\boldsymbol{W}^k=
 \begin{bmatrix}
  \text{diag}(\boldsymbol{h}_k^H)\boldsymbol{H} \\    
 f_k
\end{bmatrix} 
, \quad \boldsymbol{G}^k=
\begin{bmatrix}
  \text{diag}(\boldsymbol{v}_k^H)\boldsymbol{H} \\    
 g_k
\end{bmatrix}
\end{equation}
and the sets of TARCs can be represented as vectors and for problem transformation are defined  as $\boldsymbol{q}^{t/r}=\big[\sqrt{\beta_1^{t/r}}e^{j\phi_1^{t/r}}, \dots , \sqrt{\beta_M^{t/r}}e^{j\phi_M^{t/r}}, 1\big]^H$ for the ES and MS, and for the TS as $\boldsymbol{q}^{t/r}=\big[e^{j\phi_1^{t/r}}, \dots , e^{j\phi_M^{t/r}}, 1\big]^H$. Therefore, the optimization problem $\mathcal{P}_{1}$ can be reformulated as 
\begin{subequations}
    \begin{align}
        \begin{split}
               \mathcal{P}_{2}:\max_{\substack{\boldsymbol{Q}^r ,\boldsymbol{Q}^t,\\ \boldsymbol{\beta}^r, \boldsymbol{\beta}^t }}
                \quad & \mathlarger{\mathlarger{\sum}}_{k\in \{r, t\}}  \frac{\sigma^2 + \text{tr}(\boldsymbol{W}_B^k\boldsymbol{Q}^k)}{\sigma^2 + \text{tr}(\boldsymbol{G}_E^k\boldsymbol{Q}^k) } \Bigg / \\
                \max_{\substack{\boldsymbol{Q}^r ,\boldsymbol{Q}^t,\\ \lambda_r, \lambda_t }}
                \quad & \mathlarger{\mathlarger{\sum}}_{k\in \{r, t\}} \lambda_k  \Bigg(\frac{\sigma^2 + \text{tr}(\boldsymbol{W}_B^k\boldsymbol{Q}^k)}{\sigma^2 + \text{tr}(\boldsymbol{G}_E^k\boldsymbol{Q}^k) }\Bigg)\label{eq:10a}
        \end{split}\\
        \textrm{s.t.} \quad & \text{tr}(\boldsymbol{G}_E^k\boldsymbol{Q}^k) \ge E_k, \forall k \in \{r, t\},\label{eq:10b}\\
        \quad & \boldsymbol{Q}^k(i,i)=\boldsymbol{\beta}^k(i), \forall k \in \{r, t\}, \forall i \ne M+1, \label{eq:10c}\\
        \quad &  \boldsymbol{Q}^k(i,i)=1,\forall k \in \{r, t\}, \forall i = M+1,\label{eq:10d}\\
        \quad & \boldsymbol{Q}^k(i,i)=1,\forall k \in \{r, t\},\label{eq:10e} \forall i\\ 
        \quad & \text{rank}(\boldsymbol{Q}^k)=1, \forall k \in \{r, t\}\label{eq:10f},\\
        \quad & \boldsymbol{Q}^k \succeq 0, \forall k \in \{r, t\}, \label{eq:10g}\\
        \quad & 0 \le \beta_m^r,\beta_m^t \le 1, \beta_m^r+\beta_m^t=1, \forall m \in \mathcal{M}, \label{eq:10h} \\  
        \quad & 0 \le \lambda_r,\lambda_t \le 1, \lambda_r+\lambda_t=1, \label{eq:10i} 
    \end{align}
\end{subequations}
where $\boldsymbol{W}_B^k=\boldsymbol{W}^k(\boldsymbol{W}^k)^HP_s$, $\boldsymbol{G}_E^k=\boldsymbol{G}^k(\boldsymbol{G}^k)^HP_s$ and $\boldsymbol{Q}^k=\boldsymbol{q}^k(\boldsymbol{q}^k)^H$ and all of them are positive semi-definite matrices. Constraints \eqref{eq:10e} and \eqref{eq:10i} only valid for the TS. The optimization problem $\mathcal{P}_{2}$ is fractional programming, which is in general a non-convex optimization problem, though, it can be reformulated to an equivalent semi-definite programming (SDP) problem by utilizing the Dinkelbach transformation. It is noteworthy that due to the existence of the unit rank constraints, i.e. $\text{rank}(\boldsymbol{Q}^k)=1$ in $ \mathcal{P}_{2}$, the optimization problem is still non-convex, thus, by neglecting the rank one constraints, $ \mathcal{P}_{2}$ becomes a convex SDP problem. It is worth mentioning that if the solution of the relaxed problem is ranked one, then it is a global solution of $ \mathcal{P}_{2}$. Then to apply the SDR technique \cite{boyd2004convex}, by dropping the rank one constraints, we have
\subsubsection{Optimization for the ES protocol} For the ES protocol, the first term of objective function in $\mathcal{P}_{2}$ and constraints \eqref{eq:10b}-\eqref{eq:10d} and \eqref{eq:10g}-\eqref{eq:10h} are required. Thus, the optimization problem for the ES protocol is given as
\begin{subequations}
    \begin{align}
        \begin{split}
               \mathcal{P}_{3}:\max_{\substack{\boldsymbol{Q}^r ,\boldsymbol{Q}^t,\\ \boldsymbol{\beta}^r, \boldsymbol{\beta}^t }}
                \quad & \mathlarger{\mathlarger{\sum}}_{k\in \{r, t\}}  \sigma^2 + \text{tr}(\boldsymbol{W}_B^k\boldsymbol{Q}^k)-\\
                & \quad \quad \quad \gamma_k(\sigma^2 + \text{tr}(\boldsymbol{G}_E^k\boldsymbol{Q}^k)) \label{eq:11a}
        \end{split}\\
        \textrm{s.t.} \quad & \text{tr}(\boldsymbol{G}_E^k\boldsymbol{Q}^k) \ge E_k, \forall k \in \{r, t\},\label{eq:11b}\\
        \quad & \boldsymbol{Q}^k(i,i)=\boldsymbol{\beta}^k(i), \forall k \in \{r, t\}, \forall i \ne M+1, \label{eq:11c}\\
        \quad &  \boldsymbol{Q}^k(i,i)=1,\forall k \in \{r, t\}, \forall i = M+1,\label{eq:11d}\\
        \quad & \boldsymbol{Q}^k \succeq 0, \forall k \in \{r, t\}, \label{eq:11e}\\
        \quad & 0 \le \beta_m^r,\beta_m^t \le 1, \beta_m^r+\beta_m^t=1, \forall m \in \mathcal{M}, \label{eq:11f} 
    \end{align}
\end{subequations}
where according to the Dinkebach algorithm, $\gamma_k$ is updated in each iteration, as given in $\gamma_k^{[j+1]}= \frac{\sigma^2 + \text{tr}(\boldsymbol{W}_B^k\boldsymbol{Q}_j^k)}{\sigma^2 + \text{tr}(\boldsymbol{G}_E^k\boldsymbol{Q}_j^k)}$, where j denotes the iteration index. The algorithm terminates when the objective function of problem $\mathcal{P}_{3}$ meeting the stopping accuracy parameter $\epsilon_1$. Therefore, the optimization problem $\mathcal{P}_{3}$ is a standard convex problem and can be solved via the well-known CVX tool.
\subsubsection{Optimization for the MS protocol} Compared to ES, for the MS, the formulated optimization problem in $\mathcal{P}_{3}$ involves the additional non-convex binary constraints. Thus, as pointed in \cite{mu2021simultaneously}, we  transform the binary constraint, by using first-order Taylor expansion into an upper bound for given points $\Big\{ \beta_m^{(j)} \Big\}$ in the $j$-th iteration as follows
\begin{equation}
    \begin{split}
         \beta_m^k-(\beta_m^k)^2 &\le (\beta_m^{k(j)})^2 + (1-2\beta_m^{k(j)})\beta_m^k,\\
        & = f(\beta_m^{k(j)},\beta_m^k), \forall m \in \mathcal{M}, 
    \end{split}
\end{equation}
then we subtract the objective function of problem $\mathcal{P}_{3}$ with $\eta\sum_{m=1}^{M}\sum_{k\in \{r, t\}}f(\beta_m^{k(j)},\beta_m^k)$, where the penalty factor is gradually increased from one iteration to the next as follows: $\eta=\omega\eta$, where $\omega >0$. The algorithm for the MS terminates when the objective function and penalty term satisfy the stopping accuracy parameter $\epsilon_2$.
\subsubsection{Optimization for the TS protocol}For the TS, the optimization problem  can be rewritten as follows: 
\begin{subequations}
    \begin{align}
        \begin{split}
               \mathcal{P}_{4}:\max_{\substack{\boldsymbol{Q}^r ,\boldsymbol{Q}^t,\\ \lambda_r, \lambda_t }}
                \quad & \mathlarger{\mathlarger{\sum}}_{k\in \{r, t\}}  \lambda_k(\sigma^2 + \text{tr}(\boldsymbol{W}_B^k\boldsymbol{Q}^k)-\\
                & \quad \quad \quad \gamma_k(\sigma^2 + \text{tr}(\boldsymbol{G}_E^k\boldsymbol{Q}^k))) \label{eq:12a}
        \end{split}\\
        \textrm{s.t.} \quad & \text{tr}(\boldsymbol{G}_E^k\boldsymbol{Q}^k) \ge E_k, \forall k \in \{r, t\},\label{eq:12b}\\
        \quad & \boldsymbol{Q}^k(i,i)=1,\forall k \in \{r, t\},\label{eq:12c} \forall i\\
        \quad & \boldsymbol{Q}^k \succeq 0, \forall k \in \{r, t\}, \label{eq:12d}\\
        \quad & 0 \le \lambda_r,\lambda_t \le 1, \lambda_r+\lambda_t=1, \label{eq:12e} 
    \end{align}
\end{subequations}
after obtaining the optimal $\Big\{\boldsymbol{Q}^r ,\boldsymbol{Q}^t \Big\}$ for a given $\lambda_r$ and $\lambda_r$, the one-dimensional search method is used to obtain the optimal $\lambda_r$, $\lambda_r$ and $\Big\{\boldsymbol{Q}^r ,\boldsymbol{Q}^t \Big\}$. The algorithm terminates for the TS when the objective function of problem $\mathcal{P}_{4}$ satisfies the stopping accuracy parameter $\epsilon_1$. Thus, the optimization problem $\mathcal{P}_{4}$ is a standard convex problem and can be solved via CVX tool. If $\boldsymbol{Q}^r$ and $\boldsymbol{Q}^t$ are of rank higher than one, a ranked one optimal solution of each optimization problem can be obtained by use of a matrix decomposition theorem \cite{ai2011new}. Also, the standard Gaussian randomization method can be used to obtain an approximate solution if $\boldsymbol{Q}^r$ and $\boldsymbol{Q}^t$ are of rank higher than one.
\subsection{Complexity Analysis} 
In this section, we analyze the computational complexity of our proposed algorithms as given by $\mathcal{P}_3$, generalized $\mathcal{P}_3$ for the MS protocol and $\mathcal{P}_4$ in the SWIPT system. The complexity of the optimization problem for $\mathcal{P}_3$ and $\mathcal{P}_4$ by Dinkelbach’s method needs a polynomial time complexity, i.e., $\mathcal{O}\big[I_c\big]$, where $I_c$, denotes the number of iterations required for convergence. The complexity of generalized $\mathcal{P}_3$ for the MS protocol is $\mathcal{O}\big[I_d \times I_c\big]$, where $I_c$ and $I_d$ denote the number of iterations required for convergence and satisfaction of penalty term for binary constraints.  
\section{NUMERICAL RESULTS}
\begin{figure}[H]
    \centering
    \scalebox{0.6}
    {\includegraphics{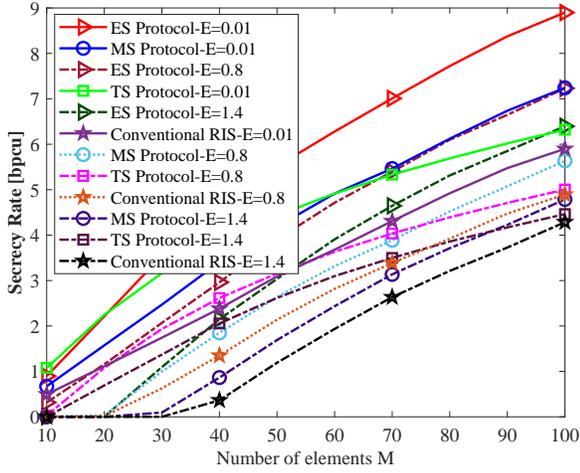}}
    \caption{Secrecy rate versus $M$ for different Eve’s energy requirement and $P_s= 20$.}
    \label{fig:2}
\end{figure}

\begin{figure}[H]
    \centering
    \scalebox{0.6}
    {\includegraphics{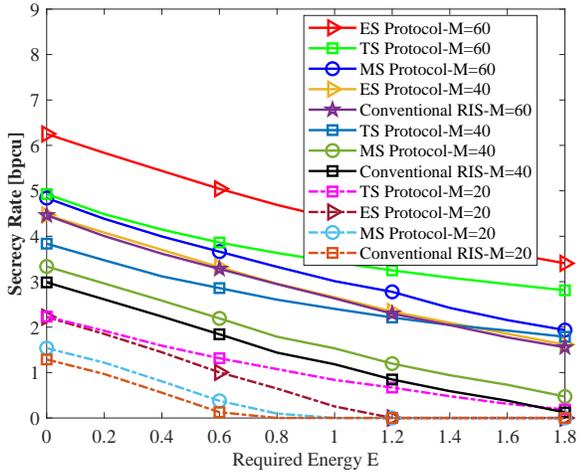}}
    \caption{Secrecy rate versus minimum energy requirement at Eve for $P_s= 20$.}
    \label{fig:3}
\end{figure}

\vfill
In this section, the performance of three protocols of STAR-RIS and conventional RIS is evaluated through simulation results. It is assumed that Alice, $\text{Bob}_r$, $\text{Bob}_t$, $\text{Eve}_r$, $\text{Eve}_t$ and STAR-RIS are respectively located at (0, 0), (12, 2), (12, -2), (10,2), (10, -2) and (8, 0) in a two dimensional area. In addition, it is assumed that $P_s=20$, $\sigma^2=1$, and for simplification and without loss of generality, each channel coefficient is modelled by a geometric path-loss and random phase, as follows; $\boldsymbol{H}=\sqrt{(1/{d_{AS}})^\alpha}\boldsymbol{q}_{AS}$, $\boldsymbol{h}_k^H=\sqrt{(1/{d_{SB_k}})^\alpha}\boldsymbol{q}_{SB_k}$ and $f_k=\sqrt{(1/{d_{AB_k}})^\alpha}{q}_{AB_k}$ where $d_{AS}$, $d_{SB_k}$ and $d_{AB_k}$ denote the relative distances for Alice-STAR-IRS, STAR-RIS-$\text{Bob}_k$ and Alice-$\text{Bob}_k$ pairs, with the path loss exponent $\alpha=2.2$. Also, $\boldsymbol{q}_{AS}=e^{j\phi_{AS}}$, $\boldsymbol{q}_{SB_k}=e^{j\phi_{SB_k}}$ and ${q}_{AB_k}=e^{j\phi_{AB_k}}$ model the random phases where $\phi_{AS}$, $\phi_{SB_k}$ and $\phi_{AB_k}$ are independent and uniformly distributed over $\big[0, 2\pi \big]$. Similar to the aforementioned geometric path loss, the wiretapping channel coefficients are specified for path loss exponent $\alpha_e=2$, which provides a stronger channel condition for the eavesdroppers.\\
Fig. \ref{fig:2} indicates the secrecy rate versus the number of STAR-RIS elements for various Eve's energy requirements. It is depicted that by increasing the number of STAR-RIS elements, the secrecy rate increases. In addition, by increasing the minimum required harvested energy at the Eves, the degradation of the secrecy rate occurs, as well, and it could be zero, for example, see the secrecy rate of cases with $E=1.4$, $M \le 20$ and $M \le 30$. Also, it can be seen that STAR-RIS protocols outperforms conventional RIS, and for sake of protocol comparison, the ES protocol outperform the MS and the TS one.\\
It is depicted in Fig. \ref{fig:3} that the secrecy rate is a non-increasing function of the minimum energy requirement at Eves. Also, by increasing the number of elements, the region of each protocol and case, enlarges.

\begin{figure}[H]
    \centering
    \scalebox{0.6}
    {\includegraphics{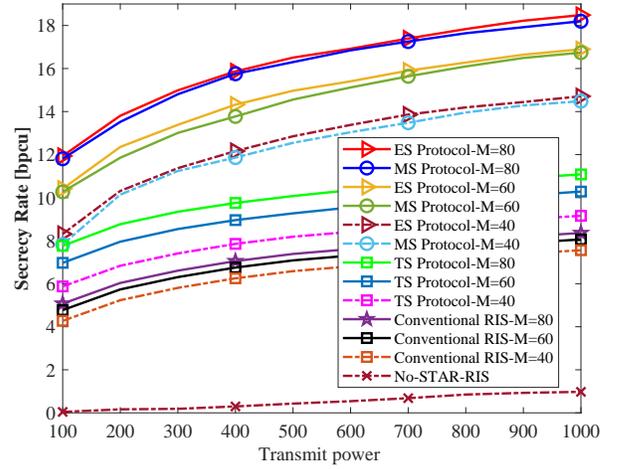}}
    \caption{Secrecy rate versus transmit power  for $E= 0.1$ with and without utilizing STAR-RIS.}

    \label{fig:4}
\end{figure}

\begin{figure}[H]
    \centering
    \scalebox{0.6}
    {\includegraphics{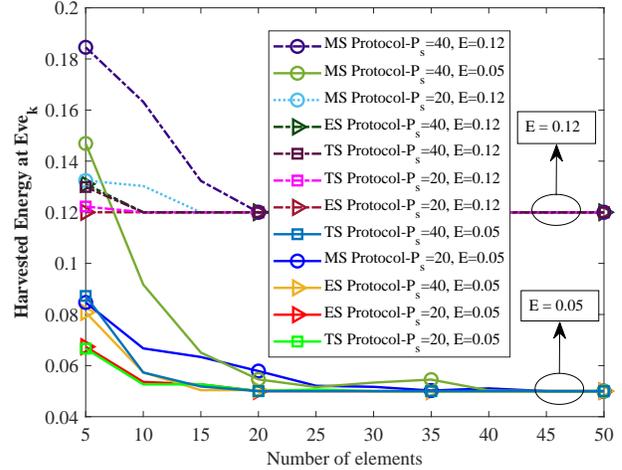}}
    \caption{Harvested energy at $\text{Eve}_k$ versus $M$ for $E \in \{0.05, 0.12 \}$ and $P_s \in \{20, 40\}$.}

    \label{fig:5}
\end{figure}

Fig. \ref{fig:4} illustrates the efficacy of increasing the transmit power for various $M$. It is depicted that by increasing $P_s$, the secrecy rate increases. Also, utilizing the STAR-RIS can dramatically improve the secrecy rate compared to that of the conventional RIS and conventional wiretap channel, i.e. without using STAR-RIS.\\
Fig. \ref{fig:5} depicts The harvested energy versus number of STAR-RIS elements for $P_s \in \{20, 40\} $, and two values of required harvested energy $E \in \{0.05, 0.12\}$. It is illustrated that by increasing $M$, due to reinforcing the degrees of freedom in the
optimization problem, the energy harvesting constraints, i.e. $P_{E_k} \ge E_k$, are gained by equality. Thereupon, for the smaller number of STAR-RIS elements, we must have $P_{E_k} > E_k$ at the cost of secrecy rate degradation.\\ \\ \\ \\ 

\section{Conclusion}
In this paper, we surveyed the performance of a system model wherein, the STAR-RIS was deployed to improve secrecy rate so long as transferring power in wireless mode to the harvesting eavesdropper nodes. In order to maximize the secrecy rate subject to harvesting minimum energy at the eavesdropper, optimizing the STAR-RIS's TARCs was pursued. We utilized the Dinkelbach transformation and SDR technique to handle the non-convexity of the main optimization problem. Eventually, through numerical results, it is indicated that due to well-optimized  STAR-RIS, the secrecy rate is improved while guaranteeing a minimum required harvested energy at the eavesdroppers, as well.
\bibliographystyle{IEEEtran}
\bibliography{STAR-RIS_SWIPT}

\begin{thebibliography}{10}
\providecommand{\url}[1]{#1}
\csname url@samestyle\endcsname
\providecommand{\newblock}{\relax}
\providecommand{\bibinfo}[2]{#2}
\providecommand{\BIBentrySTDinterwordspacing}{\spaceskip=0pt\relax}
\providecommand{\BIBentryALTinterwordstretchfactor}{4}
\providecommand{\BIBentryALTinterwordspacing}{\spaceskip=\fontdimen2\font plus
\BIBentryALTinterwordstretchfactor\fontdimen3\font minus
  \fontdimen4\font\relax}
\providecommand{\BIBforeignlanguage}[2]{{%
\expandafter\ifx\csname l@#1\endcsname\relax
\typeout{** WARNING: IEEEtran.bst: No hyphenation pattern has been}%
\typeout{** loaded for the language `#1'. Using the pattern for}%
\typeout{** the default language instead.}%
\else
\language=\csname l@#1\endcsname
\fi
#2}}
\providecommand{\BIBdecl}{\relax}
\BIBdecl

\bibitem{renzo2019smart}
M.~D. Renzo, M.~Debbah, D.-T. Phan-Huy, A.~Zappone, M.-S. Alouini, C.~Yuen,
  V.~Sciancalepore, G.~C. Alexandropoulos, J.~Hoydis, H.~Gacanin \emph{et~al.},
  ``Smart radio environments empowered by reconfigurable {AI} meta-surfaces: An
  idea whose time has come,'' \emph{EURASIP Journal on Wireless Communications
  and Networking}, vol. 2019, no.~1, pp. 1--20, 2019.

\bibitem{wu2019towards}
Q.~Wu and R.~Zhang, ``Towards smart and reconfigurable environment: Intelligent
  reflecting surface aided wireless network,'' \emph{IEEE Communications
  Magazine}, vol.~58, no.~1, pp. 106--112, 2019.

\bibitem{liu2021star}
Y.~Liu, X.~Mu, J.~Xu, R.~Schober, Y.~Hao, H.~V. Poor, and L.~Hanzo, ``{STAR}:
  Simultaneous transmission and reflection for 360° coverage by intelligent
  surfaces,'' \emph{IEEE Wireless Communications}, vol.~28, no.~6, pp.
  102--109, 2021.

\bibitem{ng2019wireless}
D.~W.~K. Ng, T.~Q. Duong, C.~Zhong, and R.~Schober, \emph{Wireless information
  and power transfer: theory and practice}.\hskip 1em plus 0.5em minus
  0.4em\relax John Wiley \& Sons, 2019.

\bibitem{abedi2018power}
M.~Abedi, H.~Masoumi, and M.~J. Emadi, ``Power splitting-based {SWIPT} systems
  with decoding cost,'' \emph{IEEE Wireless Communications Letters}, vol.~8,
  no.~2, pp. 432--435, 2018.

\bibitem{tang2019joint}
J.~Tang, Y.~Yu, M.~Liu, D.~K. So, X.~Zhang, Z.~Li, and K.-K. Wong, ``Joint
  power allocation and splitting control for {SWIPT}-enabled {NOMA} systems,''
  \emph{IEEE Transactions on Wireless Communications}, vol.~19, no.~1, pp.
  120--133, 2019.

\bibitem{masoumi2019performance}
H.~Masoumi and M.~J. Emadi, ``Performance analysis of cooperative {SWIPT}
  system: Intelligent reflecting surface versus decode-and-forward,'' \emph{AUT
  Journal of Modeling and Simulation}, vol.~51, no.~2, pp. 241--248, 2019.

\bibitem{tang2018optimization}
J.~Tang, T.~Dai, M.~Cui, X.~Y. Zhang, A.~Shojaeifard, K.-K. Wong, and Z.~Li,
  ``Optimization for maximizing sum secrecy rate in {SWIPT}-enabled {NOMA}
  systems,'' \emph{IEEE Access}, vol.~6, pp. 43\,440--43\,449, 2018.

\bibitem{saeidi2020irs}
M.~A. Saeidi and M.~J. Emadi, ``{IRS}-based secrecy rate analysis in presence
  of an energy harvesting eavesdropper,'' in \emph{2020 Iran Workshop on
  Communication and Information Theory (IWCIT)}.\hskip 1em plus 0.5em minus
  0.4em\relax IEEE, 2020, pp. 1--5.

\bibitem{shen2019secrecy}
H.~Shen, W.~Xu, S.~Gong, Z.~He, and C.~Zhao, ``Secrecy rate maximization for
  intelligent reflecting surface assisted multi-antenna communications,''
  \emph{IEEE Communications Letters}, vol.~23, no.~9, pp. 1488--1492, 2019.

\bibitem{niu2021simultaneous}
H.~Niu, Z.~Chu, F.~Zhou, and Z.~Zhu, ``Simultaneous transmission and reflection
  reconfigurable intelligent surface assisted secrecy {MISO} networks,''
  \emph{IEEE Communications Letters}, vol.~25, no.~11, pp. 3498--3502, 2021.

\bibitem{zhang2022security}
Z.~Zhang, Z.~Wang, Y.~Liu, B.~He, L.~Lv, and J.~Chen, ``Security enhancement
  for coupled phase-shift {STAR}-{RIS} networks,'' \emph{arXiv preprint
  arXiv:2208.10382}, 2022.

\bibitem{el2011network}
A.~El~Gamal and Y.-H. Kim, \emph{Network information theory}.\hskip 1em plus
  0.5em minus 0.4em\relax Cambridge university press, 2011.

\bibitem{schaible1976fractional}
S.~Schaible, ``Fractional programming. {II}, on {D}inkelbach's algorithm,''
  \emph{Management science}, vol.~22, no.~8, pp. 868--873, 1976.

\bibitem{grant2014cvx}
M.~Grant and S.~Boyd, ``Cvx: Matlab software for disciplined convex
  programming, version 2.1,'' 2014.

\bibitem{boyd2004convex}
S.~Boyd, S.~P. Boyd, and L.~Vandenberghe, \emph{Convex optimization}.\hskip 1em
  plus 0.5em minus 0.4em\relax Cambridge university press, 2004.

\bibitem{mu2021simultaneously}
X.~Mu, Y.~Liu, L.~Guo, J.~Lin, and R.~Schober, ``Simultaneously transmitting
  and reflecting ({STAR}) {RIS} aided wireless communications,'' \emph{IEEE
  Transactions on Wireless Communications}, vol.~21, no.~5, pp. 3083--3098,
  2021.

\bibitem{ai2011new}
W.~Ai, Y.~Huang, and S.~Zhang, ``New results on hermitian matrix rank-one
  decomposition,'' \emph{Mathematical programming}, vol. 128, no.~1, pp.
  253--283, 2011.

\end{thebibliography}
\end{document}